\documentclass[twocolumn,showpacs,preprintnumbers,amsmath,amssymb]{revtex4-1}

\usepackage{graphicx}
\usepackage{dcolumn}
\usepackage{bm}

\newcommand{\braket}[1]{ \langle #1 \rangle}
\newcommand{\bra}{ \langle }
\newcommand{\ket}{\rangle}

\newcommand{\bv}[1]{\mbox{\boldmath $#1$}}

\begin{document}


\title{Two-neutron halo structure of $^{\bm{31}}$F and a novel pairing anti-halo effect
}

\author{H. Masui}
\email{hgmasui@mail.kitami-it.ac.jp}
\affiliation{%
 Information Processing Center,
 Kitami Institute of Technology, Kitami, 
 090-8507, Japan
}%

\author{W. Horiuchi}
\email{whoriuchi@nucl.sci.hokudai.ac.jp}
\affiliation{
Department of Physics, Hokkaido University, Sapporo 060-0810, Japan
}

\author{M. Kimura}
\email{masaaki@nucl.sci.hokudai.ac.jp}
\affiliation{
Department of Physics, Hokkaido University, Sapporo 060-0810, Japan
}
\affiliation{
Nuclear Reaction Data Centre, Hokkaido University, Sapporo 060-0810, Japan
}
\affiliation{
Research Center for Nuclear Physics (RCNP), Osaka University, Ibaraki 567-0047, Japan
}

\date{\today}

\begin{abstract}
\begin{description}
\item[Background] A newly identified dripline nucleus
  $^{31}{\rm F}$ offers a unique opportunity  to study the two-neutron
  (2$n$) correlation at the east shore of the island of inversion 
  where the $N=28$  shell closure is lost. 
\item[Purpose] We aim to present the first three-body theoretical results
  for the radius and total reaction cross sections of $^{31}{\rm F}$.
  This will further help to investigate
  how the pairing and breakdown of the $N=28$
 shell closure influence the formation of the $2n$-halo structure
 and the anti-halo effect in this mass region. 
\item[Methods] A $^{29}{\rm F}$+$n$+$n$ three-body system is
 described by the cluster orbital shell model,
 and its total reaction cross section is calculated by the Glauber theory.
\item[Results] Our three-body calculations predict
$3.48$--$3.70$ fm for the root-mean-square radius of $^{31}{\rm F}$,
which corresponds to the total reaction cross section of
$1530$ ($1410$)--$1640$ ($1500$) mb for a carbon target
at $240$ ($900$) MeV/nucleon.
The binding mechanism and halo formation in $^{31}$F are discussed.
\item[Conclusions] The present study suggests a novel anti-halo effect in this mass
 region: When the pairing overcome the energy gap between the $p_{3/2}$ and $f_{7/2}$ orbits, the
 inversion of the occupation number of these orbits takes place, and it diminishes the $2n$-halo structure. 
\end{description}
\end{abstract}

\maketitle

{\it Introduction.--}
The neutron halo is an extreme nuclear binding mechanism in the unstable nuclei
and has been studied intensively~\cite{Ta85_PRL55,Ta13_PPNP}.
In particular, two-neutron(2$n$) halo structure of the Borromean nuclei allows us to 
investigate the competition between the mean-field and neutron correlations.
In these low-density systems, the correlations of the weakly-bound
neutrons often overcome the single-particle motion in
the nuclear mean-field~\cite{Zh93_PR231}.
As a result, the neutron pairing plays a crucial role in determining
the fundamental properties of the system.
For example, the pair potential may modify the asymptotics of the valence
neutron wave functions and considerably shrink the nuclear radius
compared to a naive estimate of the mean-field model.
This phenomenon is called the pairing anti-halo effect~\cite{Be00_PLB496,Ha11_PRC84} and is discussed for $^{32}{\rm Ne}$ in Ref.~\cite{Ha11_PRC84}. 

Experimentally, since the first observation of the $2n$-halo structure of
$^{11}{\rm Li}$~\cite{Ta85_PRL55}, several 2$n$-halo nuclei have been identified.
At present, the latest well-established heaviest 
2$n$-halo nucleus is $^{22}{\rm C}$~\cite{Ho06_PRC74,Ta10_PRL104,To16_PLB761}.
Recently, the survey has been extended to the east shore of the island
of inversion where the neutron magic numbers are lost.
Specifically, $^{31}{\rm F}$ has been identified as the dripline
nucleus of the fluorine isotopic chain~\cite{Ah19_PRL123},
and hence, it is a Borromean nucleus as $^{30}$F is unbound.
Theoretical indication of the halo structure of $^{31}$F
  has been reported very recently~\cite{Mi20_PRC101}.
  These findings further require theoretical investigations to uncover
the pairing effect on the 2$n$-halo formation in the island of inversion.  

This situation motivates us to conduct a study on the ground state of
$^{31}{\rm F}$ within a  $^{29}{\rm F}$+$n$+$n$ three-body model.
Our purpose in this study is two-fold: The first purpose is to
provide the first three-body theoretical prediction
on the 2$n$-halo structure of $^{31}{\rm F}$.
Here, we have investigated the nuclear radius expected for $^{31}{\rm F}$ 
using the Hamiltonians that reproduce the properties of neighboring nuclei.
Also, we have predicted the total reaction cross section of $^{31}$F,
which is one of the best probes for extracting the size of the unstable nuclei.
The second purpose is to unveil how the unique structure of the Fermi
surface influences the development of the 2$n$-halo structure.
The observed spin-parity and one-neutron(1$n$) halo structure
of $^{31}{\rm Ne}$~\cite{Na14_PRL112} imply
that the order of the $f_{7/2}$ and $p_{3/2}$ orbits is inverted.
Namely, the $N=28$ shell closure appears to be melted in this mass region.
It is noted that the neutron halo appears only in the orbits
with $\ell=0$ and 1~\cite{Ri92_NPA548}.
Therefore, the $p_{3/2}$ orbit can be a halo orbit
but the $f_{7/2}$ orbit cannot be.
Since the Fermi surface of $^{31}{\rm F}$ consists of these two orbits,
it is of interest and importance to investigate how the order of these
orbits and the energy gap affect the 2$n$-halo structure.

{\it Theoretical model.--}
To address these questions, we employ
the cluster-orbital shell model (COSM)~\cite{Su88_PRC38}
to describe the $^{29}{\rm F}$+$n$+$n$ three-body system,
in which the Hamiltonian consists of one- and two-body terms as,
\begin{equation}
  \label{eq:COSM_01}
  \hat{H} = \sum_{i=1}^{2}
  (\hat{T}_{i} + \hat{V}_{i} + \lambda \hat{\Lambda}_{i})
  +\hat{t}_{12} + \hat{v}_{12},
\end{equation}
where $\hat{T}_{i}$, $\hat{V}_{i}$, and $\hat{v}_{12}$ are the kinetic
energy of the valence neutron, potential between the $^{29}{\rm F}$ core
and a valence neutron, and potential between the valence neutrons, respectively.
The center-of-mass kinetic energy is properly subtracted,
which induces the recoil term $\hat{t}_{12} =\bv{p}_{1} \cdot \bv{p}_{2} /M$ 
with $M$ and $\bm{p}_i$ being the mass of the $^{29}{\rm F}$ core
and the momentum operator of the $i$th valence neutron.
$\lambda \hat{\Lambda}_{i}$
is the pseudo potential for eliminating the Pauli forbidden states (PFS),
\begin{align}
 \lambda\hat{\Lambda}_i=\lambda\sum_{\beta\in \rm PFS}|\varphi_\beta(i)\ket\bra\varphi_\beta(i)|,
\end{align}
where ${\varphi}_{\beta}(i)$ denotes the single particle states occupied by the neutrons in 
$^{29}{\rm F}$, i.e., PFS. By using sufficiently large value of $\lambda$, they are
variationally removed~\cite{Kukulin78}.  

In this study, we assume an inert and spinless $^{29}{\rm F}$ core with the $N=20$ shell closure,
and hence, PFS corresponds to the $0s_{1/2}$, $0p_{3/2}$, $0p_{1/2}$, $0d_{5/2}$, $1s_{1/2}$ and $0d_{3/2}$ orbits.
We note that this assumption might be too strong
because $^{29}{\rm F}$ is located
in the island of inversion~\cite{Jenny16,Peter17} and also
a shell model calculation \cite{Caurier14} predicts
that the $0\hbar\omega$  component amounts to 60\% and
the remaining 40\% are from the neutron excitations across $N=20$ shell gap.
However, for the sake of simplicity,  in the present study
we use this closed shell assumption for the $^{29}$F configuration.

With this assumption, the antisymmetrized basis function for the
$^{29}{\rm F}$+$n$+$n$ system with the spin-parity of $0^{+}$ is given as
\begin{eqnarray}
  \label{eq:Basis_001}
 \Phi_{pq\ell j}  &  \equiv &  {\mathcal A}
  \left\{ [
  \phi_{p\ell j}(1) \otimes\phi_{q\ell j}(2)
            ]_{00}
  \right\},
\end{eqnarray}
where $\mathcal{A}$ is the antisymmetrizer. The single-neutron basis $\phi_{p\ell j}$  
has the Gaussian form, 
\begin{equation}
  \label{eq:COSM_06}
 \phi_{p\ell j}(\bv{r})
 = r^{\ell}\exp(-r^2/(2b_p^2))
[Y_{\ell} \otimes \chi_{1/2}]_{j m},
\end{equation}
which is flexible enough to describe the correct asymptotic behavior
and di-neutron correlation~\cite{My14_PPNP79,Ma14_PRC89}.
The orbital angular momentum $\ell$ is taken up to $5$,  
and $20$ range parameters with the geometric progression are adopted:
$b_{p} = 0.1 \times 1.25^{p-1}$ fm $(p=1,\dots, 20)$.   
The ground state of the $^{31}{\rm F}$ is described as a sum of the basis wave functions,
\begin{align}
 \Psi = \sum_{pq\ell j} c_{pq\ell j}\Phi_{pq\ell j},
\end{align}
where the coefficients $c_{pq\ell j}$ are determined by the diagonalization of the Hamiltonian.

{\it Potential setup.--}
The $^{29}$F-$n$ potential consists of the central $V_{c}$
and spin-orbit  $V_{\ell s}$ terms with the Woods-Saxon form
\begin{align}
  V_{\rm c}(r)&=-V_{0}f(r),\quad
  V_{\ell s}(r)=V_{1}r_{0}^{2}\bm{\ell}\cdot \bm{s}\frac{1}{r}\frac{d}{dr}f(r),\label{eq:pot}
\end{align}
where $f(r)=\left\{1+\exp{\left[(r-R)/a\right]}\right\}^{-1}$ 
with the radius parameter $R=r_0A^{1/3}$.
The parameters of these terms are usually determined
so as to reproduce the properties of $^{30}$F.
However, no experimental information is available other than the fact that 
$^{30}{\rm F}$ is unbound. Therefore, we employ the parameters as for the $^{30}{\rm Ne}$-$n$ potential used in the analysis of
the $1n$-halo nucleus $^{31}{\rm Ne}$,
which is located next to $^{31}{\rm F}$ in the nuclear chart.
In Ref.~\cite{Ho10_PRC81}, six different parameter sets were proposed:
Some of them locate the $f_{7/2}$ orbit below the $p_{3/2}$ orbit (normal order),
while the others yield the $p_{3/2}$ orbit lower
than the $f_{7/2}$ orbit (inverted order).
As a result, they yield notable differences in the total reaction cross sections of $^{31}{\rm Ne}$,
and the experimental data supports the inverted order of the $f_{7/2}$ and $p_{3/2}$ orbits.

\begin{table*}[thb]
  \caption{Parameters for the $^{29}{\rm F}$-$n$ potential and the results
    for $^{31}{\rm F}$ obtained by the three-body calculations. 
    The spin-orbit coupling strength is taken
    as $V_{1} =  22- 14[(N-Z)/ A]$.
    $\varepsilon(p)$ and $\varepsilon(f)$ denote the
 $p_{3/2}$ and $f_{7/2}$ resonance energies of the two-body system ($^{29}{\rm F}+n)$. $S_{2n}$,
 $\sqrt{r_m^2}$, $N(p)$ and $N(f)$ are the two-neutron separation energy, rms matter radius, and valence neutron occupation number of the $p_{3/2}$ and $f_{7/2}$ orbits in
 the $^{31}{\rm F}$, respectively. The rows shown by bold face (7th and 12th rows) display the
 parameters that yield the largest (case A) and smallest (case B) radii of the $^{31}{\rm F}$,
 respectively.}\label{tab:result}  
\begin{ruledtabular} 
\begin{tabular}{ccccccccccc}
& \multicolumn{3}{c}{parameter}&\multicolumn{3}{c}{$^{29}{\rm F}$+$n$}&\multicolumn{4}{c}{$^{29}{\rm F}$+$n$+$n$}\\
\cline{2-4}\cline{5-7}\cline{8-11}
& $r_0 $[fm] & $a$ [fm] & $V_0$ [MeV] & $\varepsilon(p)$ [MeV] & $\varepsilon(f)$ [MeV] &
$\Delta\varepsilon$ [MeV] & $S_{2n}$ [MeV] & $\sqrt{\braket{r^2_m}}$ [fm] &$N(p)$ & $N(f)$
\\\hline
&1.20 & 0.65 & 52.0 & 0.03 & 0.41 & 0.38 & 1.10 & 3.54 & 1.10 & 0.75  \\
&     &      & 51.0 & 0.17 & 0.84 & 0.67 & 0.52 & 3.61 & 1.40 & 0.43  \\
&     & 0.70 & 51.0 & 0.02 & 0.89 & 0.87 & 0.81 & 3.61 & 1.58 & 0.24  \\
&     &      & 50.0 & 0.15 & 1.28 & 1.13 & 0.44 & 3.67 & 1.65 & 0.16  \\
&     & 0.75 & 50.0 & 0.01 & 1.25 & 1.24 & 0.85 & 3.64 & 1.70 & 0.12  \\
&     &      & 49.0 & 0.14 & 1.61 & 1.47 & 0.41 & 3.70 & 1.72 & 0.09\rule[-10pt]{0pt}{0pt} \\
\bf case B&\bf 1.25 & \bf 0.65 & \bf 48.0 & \bf 0.06 & \bf 0.13 & \bf 0.08 & \bf 1.37 & \bf 3.48 & \bf 0.38 & \bf 1.53  \\
&     &      & 47.0 & 0.18 & 0.57 & 0.39 & 0.61 & 3.53 & 0.71 & 1.16  \\
&     & 0.70 & 47.5 & 0.01 & 0.42 & 0.41 & 1.12 & 3.55 & 1.03 & 0.82  \\
&     &      & 46.5 & 0.14 & 0.83 & 0.69 & 0.53 & 3.63 & 1.35 & 0.48  \\
&     & 0.75 & 46.5 & 0.01 & 0.84 & 0.83 & 0.86 & 3.63 & 1.54 & 0.30  \\
\bf case A& &      & \bf 45.5 & \bf 0.14 & \bf 1.22 & \bf 1.09 & \bf 0.38 & \bf 3.70 & \bf 1.63 & \bf 0.19  \\
\end{tabular}
\end{ruledtabular}
\end{table*}

When we apply the same potentials to the $^{29}{\rm F}$+$n$ system taking into account the mass
dependence of the parameters, we find that in all cases
the $p_{3/2}$ level is lower than the $f_{7/2}$ one.
Namely, the magic number $N=28$ is broken in all the cases due to 
the weaker core-$n$ attraction.
However, we also found that most sets (five of the six parameter sets)
bound $^{30}{\rm F}$ which contradict the observation.
Therefore,  we slightly weakened the potential strength $V_{0}$
so that $^{30}{\rm F}$ is unbound but $^{31}{\rm Ne}$ is bound.
The fixed parameter sets are listed in Table~\ref{tab:result}.
We prepare two variations of $V_0$ for each of the original
six parameter sets, and in total 12 sets are generated.
From these variations, one gives a very weak binding for the $p_{3/2}$ orbit
(less than $100$ keV), while the other binds more strongly.
It is also noted that all the parameter sets
locate the $f_{7/2}$ orbit above the $p_{3/2}$ orbit, and the energy gap between them, 
$\Delta \varepsilon=\varepsilon(f)-\varepsilon(p)$, ranges from $0.08$ to $1.47$ MeV.  

For the interaction between the valence neutrons $\hat{v}_{12}$,
we use the Minnesota
potential~\cite{Th77_NPA286} with the exchange parameter $u=1.0$.
It is noted that the Minnesota potential combined with
the $^{30}{\rm Ne}$-$n$ potentials given in Ref.~\cite{Ho10_PRC81}
yield reasonable binding energies for the $^{32}{\rm Ne}$,
between $1.8$ and $2.3$ MeV.
Therefore, we use the Minnesota potential without any modifications.

{\it Results.--}
Using the parameter sets listed in Table~\ref{tab:result},
we perform the three-body calculations for the  $^{29}{\rm F}$+$n$+$n$ system.
The numerical results are listed in Table~\ref{tab:result}.
Though the two-neutron separation energy, $S_{2n}$,
depends mainly on the strength of
the mean-field potential $V_{0}$,
we find that the variation of $S_{2n}$ is small
from $0.41$ to $1.37$ MeV
within the parameter sets employed in this paper.
On the other hand, the root-mean-square (rms) radius of
the matter and the valence neutron density distributions
strongly depend on the choice of the parameter set.

\begin{figure}[h!tbp]
 \begin{center}
  \includegraphics[width=0.9\hsize]{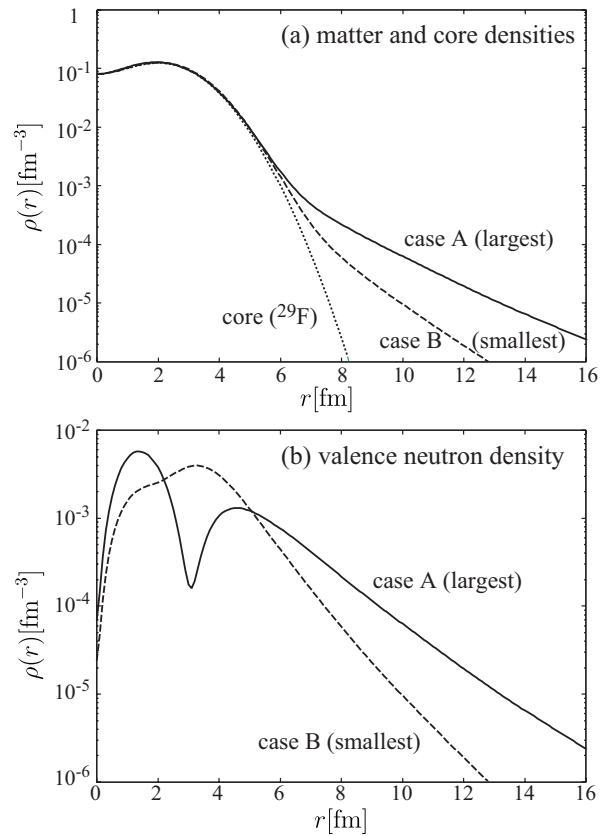}
  \caption{Panel (a): Density distributions shown by
    solid and dashed lines denote the total matter densities
    of $^{31}{\rm F}$ for the largest ($\sqrt{\braket{r_m^2}}=$3.70 fm)
    and smallest (3.48 fm) radii cases
    normalized to the mass numbers, respectively.
    The core ($^{29}{\rm F}$) density distribution is drawn in a dotted line.   
  Panel (b): Valence neutron density distributions in the largest and smallest radii cases. }
  \label{fig:density1} 
 \end{center}
\end{figure}

For more clarity, Fig.~\ref{fig:density1} displays the matter density distributions for the largest
radius case ($3.70$ fm, case A in Table~\ref{tab:result})
and for the smallest radius case ($3.48$ fm, case B).
Here, the density distributions are calculated as follows.
First, the density of $^{30}{\rm Ne}$ is calculated assuming
the harmonic oscillator wave function with the $N=20$ shell closure.
The oscillator width is chosen so as to reproduce the total reaction cross section of the
$^{30}$Ne on a carbon target at 240 MeV~\cite{Ta12_PLB707}.
The density of the core nucleus $^{29}$F is constructed
by removing a proton from the $^{30}$Ne density.
With this procedure, the rms radius of the $^{29}$F is calculated as $3.37$ fm.
Then, the total matter density distribution of the $^{31}{\rm F}$
is obtained as a sum of the core density and the valence neutron density
calculated by the three-body model.
The center-of-mass correction is ignored because the recoil effect
is expected to be small due to a large mass number of the core.

As observed in Fig.~\ref{fig:density1} (a),
the largest radius case (case A) exhibits a typical halo structure.
The rms radius is enlarged by $0.33$ fm compared to the core nucleus,
which is comparable with the case of $^{22}$C
($\sim 0.4$ fm enlargement)~\cite{To16_PLB761,Na18_PRC97}.
On the other hand, the enhancement is rather small ($0.11$ fm)
for the smallest radius case.
This difference originates from the asymptotics of the valence neutron
density as observed in Fig.~\ref{fig:density1} (b).
The largest radius case has a dip around $r=3$ fm and a long asymptotic tail,
while the smallest radius case does not.
This apparently indicates that the largest radius case is dominated
by the valence neutrons in the $1p_{3/2}$ orbit,
which have a node and an extended asymptotic wave function
due to the small centrifugal barrier.
This is consistent with the valence
neutron occupation number listed in Table~\ref{tab:result}.
Here the occupation numbers are normalized to two, and hence, $N(p)+N(f) \leq 2$ holds.
This demonstrates the dominance of the $(p_{3/2})^2$ configuration in case A.
It is interesting to note that the $(f_{7/2})^2$ configuration is dominant in case B,
even though the $f_{7/2}$ orbit is located higher than the $p_{3/2}$ orbit.
As a general tendency for all parameter sets, we can see that
the larger the occupation of the $p_{3/2}$ orbit, the larger the radius. 

The present calculations predict a value of $0.44$--$1.37$ MeV for $S_{2n}$, and
$3.48$--$3.70$ fm for the rms matter radius of $^{31}$F.
The halo formation in $^{31}$F strongly depends on the occupation of the $p_{3/2}$ orbit.
We note that a similar behavior is also seen
in the recent three-body analysis of $^{29}$F~\cite{Si20_PRC}. 

The total reaction or interaction cross sections at high incident
energies may be the best probe to study the matter radius of $^{31}{\rm F}$.
Since the first discovery of the halo nucleus $^{11}$Li~\cite{Ta85_PRL55},
they have been used as a standard and direct way to extract the size
properties of the unstable nuclei~\cite{Oz01_NPA693,Ta13_PPNP}.
To predict the total reaction cross sections, here we employ
the nucleon-target profile function in the Glauber
theory~\cite{Glauber} (NTG~\cite{NTG})
which only requires the nuclear density distributions and
nucleon-nucleon profile functions.
With an appropriate choice for a set of the nucleon-nucleon profile
functions~\cite{Ho07_PRC75,Ab08_PRC77}, the NTG offers a nice reproduction of the total
reaction cross section data, including the neutron-rich unstable
nuclei, without introducing any adjustable parameters~\cite{Ho07_PRC75,Ho12_PRC86,Ho15_proc}.  

The total reaction cross sections of $^{31}$F on a carbon target calculated from the density profile of
the largest (smallest) radius case  are obtained as $1530$ ($1410$) and $1640$ ($1500$) mb at the
incident energies, $240$ and $900$ MeV/nucleon, respectively.
These values may be compared with those for $^{29}{\rm F}$ calculated
as $1340$ and $1410$ mb for $240$ and $900$ MeV/nucleon, respectively.
As a result, about $15$\% increase in the cross section
from  $^{29}{\rm F}$ to $^{31}{\rm F}$ is obtained for the largest
radius case, and about $5$\% increase for the smallest radius case.
It is noted that the experimental uncertainties on the carbon
target amount to a few per cent~\cite{Ka11_PRC83, Ka11_PRC84, Ta12_PLB707, Ta14_proc}, and hence,
these differences are significant enough to distinguish between the two cases.
We also predict that  the total reaction cross sections
on a proton target for the largest (smallest) radius case are 
$551$ ($533$) and $574$ ($556$) mb at $240$ and $900$ MeV/nucleon, respectively.

{\it Discussions.--}
\begin{figure}[h!tbp]
 \includegraphics[width=0.9\hsize]{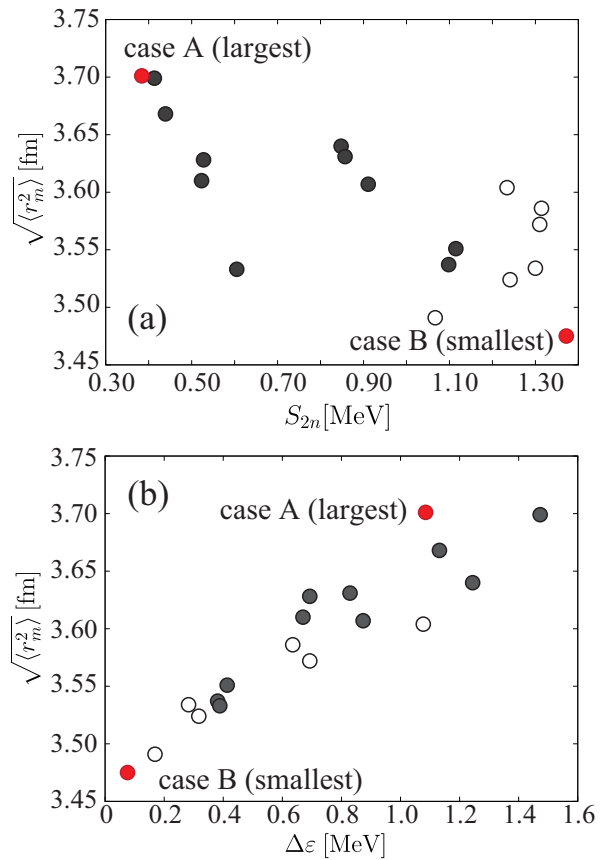}
 \caption{(color online) Panel (a): Rms radius versus $S_{2n}$ of $^{31}{\rm F}$ obtained  by different
 potential parameter sets. Filled circles show the data listed in Table~\ref{tab:result}, red
 circles show the largest and smallest radius cases, and open circles are the results obtained by using the original parameter sets~\cite{Ho10_PRC81}
 for the sake of comparison.
 Panel (b): Similar to the panel (a), but here the data is for the rms radius versus the energy
 gap $\Delta \varepsilon$. }
  \label{fig:s2n-r}
\end{figure}
\begin{figure}[h!tbp]
 \includegraphics[width=0.9\hsize]{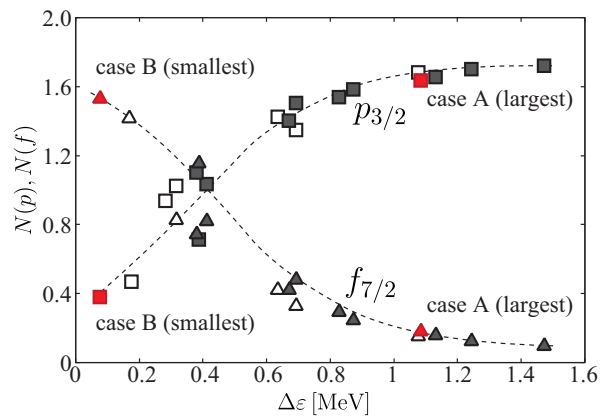}
 \caption{(color online) Valence neutron occupation numbers in the $p_{3/2}$ (boxes) and 
 $f_{7/2}$ (triangles) orbits as a function of the energy gap. 
 Filled symbols show the data listed in Table~\ref{tab:result},
 red symbols are the largest and smallest radius cases, and open symbols are the results obtained by using the original parameter sets~\cite{Ho10_PRC81}
 for the sake of comparison. }
 \label{fig:occ} 
\end{figure}
An important question to be addressed is what mechanism determines
the $2n$-halo formation in $^{31}{\rm F}$.
In the case of the $1n$-halo nuclei, the one neutron separation energy strongly
correlates with the nuclear radius, since it determines
the asymptotics of the valence neutron wave function.
However, in the present three-body system, we do not find a strong correlation
between $S_{2n}$ and $\sqrt{\braket{r_m^2}}$,
although there is a trend that the radius decreases as $S_{2n}$ increases (Fig.~\ref{fig:s2n-r} (a)).
The data points are broadly scattered,  and the correlation is not very strong.
For example, we can pick up a data point that gives a small
separation energy of $S_{2n}=0.61$ MeV, but it actually gives a small radius of 3.53 fm.    

On the other hand, we find a strong correlation between the radius and energy gap
$\Delta\varepsilon$ as shown in Fig.~\ref{fig:s2n-r} (b).
The correlation between two variables may be quantified
by the Pearson correlation coefficient (PCC),
\begin{align}
  r_{xy} = \frac{
   \sum_{i = 1}^{M} (x_i - \overline{x})
  (y_i - \overline{y})}
  { \{ \sum_{i = 1}^{M} 
  (x_i - \overline{x})^{2} \}^{1/2} \, 
  \{ \sum_{i = 1}^{M} 
(y_i - \overline{y})^{2} \}^{1/2} } \mbox{ ,}
\end{align}
where $M$ is the total number of data points, and $\overline{x}$ and $\overline{y}$ are the mean
values of the variables $x$ and $y$, respectively.
By definition, $r_{xy} $ has a value ranging from $-1$ to $1$
where the sign represents the positive- or negative-correlation.
When the two variables $x$ and $y$ have a strong linear correlation,
$r_{xy}$ approaches $\pm 1$.
The calculated PCC between the radius and $S_{2n}$ is $-0.73$, which is a weak correlation.
On the contrary, the PCC between the radius and energy gap is $0.93$,
indicating a strong correlation between them.

A strong correlation is also found between the energy gap
$\Delta \varepsilon$ and valence neutron occupation numbers $N(p)$ and $N(f)$.
Figure~\ref{fig:occ} demonstrates that the occupation number is
insensitive to the energy of the $p_{3/2}$ orbit (weak binding of $p_{3/2}$),
but depends only on the energy gap $\Delta \varepsilon$.
It should be noted that the occupation number of the $f_{7/2}$ orbit
becomes larger than that of the $p_{3/2}$ orbit at smaller energy gaps,
although the $f_{7/2}$ orbit is always located at higher energy than the $p_{3/2}$ orbit.  
The reason for this is qualitatively understood by a two-level pairing model~\cite{Li65_NP62}.
Assuming the constant pairing interaction, it can be shown that
the neutron pair tends to occupy the orbit with larger degeneracy
as the energy gap between the two orbits becomes smaller.
In the present case, we observe the inversion of the occupation number
taking place at $\Delta\varepsilon=0.40$ MeV.    

From these considerations,  the mechanism for the formation
and suppression of the halo structure in $^{31}{\rm F}$
is summarized as follows: In this mass region, the $p_{3/2}$ resonance is likely
located below the $f_{7/2}$ resonance.
The $^{29}{\rm F}$+$n$ system is unbound, but the $^{29}{\rm F}$+$n$+$n$
three-body system is bound with the help of the pairing correlation.
When the energy gap between the $p_{3/2}$ and $f_{7/2}$ orbits is large enough,
the valence neutrons predominantly occupy the $p_{3/2}$ orbit and form the 2$n$-halo structure.
On the other hand, when the energy gap is small, the valence neutrons occupy the
$f_{7/2}$ orbit, which has a larger degeneracy to gain a larger pairing energy.
As a result, even though the $f_{7/2}$ orbit is located above the $p_{3/2}$
orbit, the inversion of the occupation numbers takes place,
and the halo structure disappears.
Thus, the pairing correlation binds the $^{29}{\rm F}$+$n$+$n$, 
but it diminishes the 2$n$-halo if the energy gap is too small, or, 
in other words, if the breaking of the $N=28$ magic number is not strong enough. 
Thus, the formation and suppression of the $2n$-halo structure of $^{31}$F is determined by 
a delicate balance between the energy gap of the single-neutron orbits and pairing interaction.   

To our knowledge, this suggests an interesting and unexplored pairing effect on the halo structure  
and is regarded as a novel pairing anti-halo effect. 
At present, none of the quantitative information for the ingredient of this novel phenomenon is
available: The resonance parameters of the $p_{3/2}$ and $f_{7/2}$ in $^{30}$F,  their energy gap,
and the two-neutron separation energy and rms matter radius of $^{31}{\rm  F}$.
These experimental data are crucially important for confirming
the $2n$-halo structure of $^{31}{\rm  F}$
and for establishing the novel pairing anti-halo effect.

 As we noted that the results presented in this paper
  are based on the assumption that the neutron magicity of $^{29}$F
  is not broken. To be more realistic, we need to consider
  holes of the $^{29}$F core configuration 
  in which the $d_{3/2}$ orbit has to be considered in
  the three-body calculation as was investigated
  in the $^{27}{\rm F}+n+n$ three-body model~\cite{Si20_PRC}.
  Since the occupancy of the $d_{3/2}$
  also play a role to suppress the rms radius,
  this novel anti-halo effect can occur depending on
  the shell gap between $d_{3/2}$ and $p_{3/2}$ orbits.

{\it Summary.--}
In summary, we have studied the $2n$-halo structure of the neutron dripline nucleus 
$^{31}{\rm F}$. Three-body ($^{29}{\rm F}$+$n$+$n$) model
calculations were conducted using $12$ different parameter sets
for the $^{29}{\rm F}$-$n$ potential,
which do not contradict to scarce experimental information.
From the calculated matter density distributions,
the Glauber model analysis was also performed to predict
the total reaction cross sections
of $^{31}{\rm F}$ on carbon and proton targets. 

We found that the two-neutron separation energy does not strongly depend
on the choice of the
potential parameter sets, but the rms radius does. The large
variation in the rms radii, ranging from $3.48$ to $3.70$ fm,
originates from the formation and suppression of 
the $2n$-halo structure depending on the choice
of the $^{29}{\rm F}$-$n$ potentials.
We predict that the variation of the radius will be reflected
in the total reaction cross sections as 5 to 15\% increase from
$^{29}{\rm F}$ to $^{31}{\rm F}$, which is large enough to be distinguished experimentally.

Behind the formation and suppression of the $2n$-halo structure,
we found a novel pairing effect.
As demonstrated, the $p_{3/2}$ orbit is always located below
the $f_{7/2}$ orbit for any choice in the parameter sets,
and the magnitude of the energy gap between the two orbits determines
the $2n$-halo formation and suppression.
When the energy gap is large, $\Delta \varepsilon \gtrsim 0.4$ MeV,
the valence neutrons predominantly occupy the $p_{3/2}$ orbit,
whose extended asymptotic wave function forms the halo structure.
On the contrary, when the energy gap is small,
$\Delta \varepsilon \lesssim 0.4$ MeV, the valence neutrons
are promoted to the $f_{7/2}$ orbit to gain a larger pairing energy,
and as a result,  the halo structure disappears.
In other words, when the pairing correlation overcomes the
single-particle energy gap, it diminishes the $2n$-halo structure.
This provides a new insight into the role of the pairing
in the dripline nuclei.   

Finally, it is emphasized that the experimental data for the $p_{3/2}$ and $f_{7/2}$ resonances of
$^{30}$F, two-neutron separation energy, and total reaction cross section of
$^{31}{\rm F}$ are indispensable to establish this novel pairing anti-halo effect.


\acknowledgments
We thank J. Singh for a careful reading of the manuscript.
This work was in part supported by JSPS KAKENHI
Grant Nos. 18K03636, 18K03635, 18H04569, 19H05140, and
19K03859, and the collaborative research program 2019,
information initiative center, Hokkaido University.

\end{document}